\documentstyle[12pt]{article}
\begin{document}
\begin{titlepage}
\begin{center}
\hfill YITP-SB-03/32 \\
\vskip 20mm

{\Huge Four dimensional supersymmetrization of ${\cal R}^4$}
\footnote{Talk given at the international conference on "20 Years of SUGRA and 
Search for SUSY and Unification", Northeastern University, Boston, March 17-20
2003.}
\vskip 10mm
Filipe Moura
\vskip 4mm
{\em C. N. Yang Institute for Theoretical Physics \\
State University of New York \\
Stony Brook, NY 11794-3840, U.S.A}\\
{\tt fmoura@insti.physics.sunysb.edu}
\vskip 6mm
\end{center}
\vskip .2in

\begin{center} {\bf Abstract } \end{center}
\begin{quotation}
\noindent
We review our recent works on the supersymmetrization of the leading string 
correction (the ${\cal R}^4$ term) to ${\cal N}=1,2$ supergravity theories in
four dimensions. We show that, in the "old minimal" formulations of these 
theo\-ries, when going on-shell in the presence of this correction, the 
auxi\-liary fields which come from multiplets with physical fields cannot be 
eliminated, but those ones that come from compensating multiplets without any 
physical fields can be eliminated. We conjecture similar results for other 
versions of these theories.
\end{quotation}
\vfill
\flushleft{\today}
\end{titlepage}
\eject


\section{${\cal N}=1,2$ supergravity: from conformal to Poincar\'e}
\indent

${\cal N}=1,2$ Poincar\'e supergravities can be obtained from the 
corresponding conformal theories by consistent couplings to compensating 
multiplets that break superconformal invariance and local U($\cal N$). There 
are different possible choices of compensating multiplets, leading to different
formulations of the Poincar\'e theory. What is special about these theories 
is the existence of a completely off-shell formalism. This means that, for each
of these theories, a complete set of auxiliary fields is known (actually, there
exist three known choices for each theory). In superspace this means that, 
after imposing constraints on the torsions, we can completely solve the Bianchi
identities without using the field equations \cite{gwz79,howe82}, and there is
a perfect identification between the superspace and $x$-space descriptions. 

\subsection{The ${\cal N}=1$ case}
\indent

The fields of ${\cal N}=1$ conformal supergravity multiplet are the graviton 
$e_\mu^m$, the gravitino $\psi_\mu^A$ and a U(1) gauge field $A_\mu$. In order 
to break the superconformal invariance and obtain the "old minimal" formulation
of ${\cal N}=1$ Poincar\'e supergravity \cite{sw78,fvn78}, one must impose some
constraint which restricts a linear combination of the superconformal 
parameters 
$L, \overline{L}$ to a compensating chiral multiplet \cite{ht78}. This is 
achieved by setting the nonconformal superspace torsion constraint
\begin{equation}
T_{Am}^{\ \ \ m}=0
\end{equation}
This formulation of supergravity is described in terms of the superfields $R$, 
$G_m$, $W_{ABC}$, their complex conjugates and their covariant derivatives. 
$R$ and $W_{\dot A\dot B\dot C}$ are antichiral:
\begin{equation}
\nabla^A R=0, \, \nabla_A W_{\dot A\dot B\dot C}=0
\end{equation}
The Bianchi identities and all the torsion constraints imply the following 
off-shell differential relations between the ${\cal N}=1$ supergravity 
superfields:
\begin{eqnarray}
\nabla^A G_{A\dot B} &=& \frac{1}{24} \nabla_{\dot B} R \label{diffg} \\
\nabla^A W_{ABC} &=& i \left( \nabla_{B \dot A} G_C^{\ \ \dot A} +\nabla_{C
\dot A} G_B^{\ \ \dot A} \right) \label{diffw}
\end{eqnarray}
which imply the relation
\begin{equation}
\nabla^2 \overline{R} - \overline{\nabla}^2 R = 96 i \nabla^n G_n 
\label{wb203}
\end{equation}
At $\theta=0$, we have $\left. G_m \right|=A_m$, now an auxiliary field. The 
(anti)chirality condition on $R, \overline{R}$ implies their $\theta=0$ 
components (resp. the auxiliary fields $M-iN, M+iN$) lie in antichiral/chiral 
multiplets (the compensating multiplets); (\ref{diffg}) shows the spin-1/2 
parts of the gravitino lie on the same multiplets (because $\left. \nabla_A 
G_{B \dot B} \right|$ includes the gravitino curl) and, according to 
(\ref{wb203}), so does $\partial^\mu A_\mu$.

The superspace action for this formulation of supergravity is given by
\cite{wz781}:
\begin{equation}
{\cal L}_{SG}=\frac{1}{2 \kappa^2} \int E d^4\theta, \, 
E=\mbox{sdet}E_\Lambda^{\ \ M} \label{pure}
\end{equation}

\subsection{The ${\cal N}=2$ case}
\indent

The ${\cal N}=2$ Weyl multiplet has 24+24 degrees of freedom. Its field 
content is given by the graviton $e_{\mu}^m$, the gravitinos $\psi_\mu^{A a}$,
the U(2) connection $\widetilde{\Phi}_\mu^{ab}$, an antisymmetric tensor 
$W_{A \dot A B \dot B} = 2 
\varepsilon_{\dot A \dot B} W_{AB} +2 \varepsilon_{AB} W_{\dot A \dot B}$, a 
spinor $\Lambda_A^a$ and, as auxiliary field, a dimension 2 scalar $I$. 

In U(2) ${\cal N}=2$ superspace there is an off-shell solution to the Bianchi
identities. The torsions and curvatures can be expressed in terms of 
superfields $Y_{AB}$, $U_{A \dot A}^{ab}$, $X_{ab}$ and the previously 
mentioned $W_{AB}$, their complex conjugates and their covariant derivatives. 
In order to obtain the Poincar\'e theory, the first step is to couple to the 
conformal theory an abelian vector multiplet (with central charge), which 
includes a vector $A_\mu$ described, in superspace, by a 1-form $A_\Pi$. The 
field strength $F_{\Pi \Sigma}$ satisfies its own Bianchi identities 
$\nabla_{\left[\Gamma \right.} F_{\left.\Pi \Sigma \right \}}=0$. After 
imposing conventional and conformal-breaking constraints in some of its 
components, we solve the Bianchi identities for $F_{\Pi \Sigma}$ in terms of 
all the previous superfields. $W_{mn}$ is now related to the vector field 
strength $F_{mn}$, while $Y_{mn}$, $X_{ab}$, $U_m$ emerge as auxiliary fields,
$U_\mu$ being the U(1) piece of $\widetilde{\Phi}_\mu^{ab}$. 
Although the algebra closes with this multiplet, it does not admit a consistent
lagrangian because of the higher-dimensional scalar $I$.

In order to obtain the "old minimal" Poincar\'e theory, we break the remaining 
local SU(2) invariance, by restricting its parameter $L^{ab}$ to a compensating
nonlinear multiplet \cite{whp80}. This is achieved by imposing the following 
constraint on the fermionic SU(2) connection:
\begin{equation}
\Phi_A^{abc}=2 \varepsilon^{a \underline{b}} \rho_A^{\underline{c}} 
\label{defrho}
\end{equation}
This constraint requires introducing a new fermionic superfield $\rho_A^a$. We
also introduce its fermionic derivatives $P$ (a complex scalar) and $H_m$. The 
previous SU(2) connection $\Phi_\mu^{ab}$
is now an unconstrained auxiliary field. The divergence of the vector field 
$H_m$ is constrained by  
\begin{eqnarray}
I &=& 4 R -6 \nabla_{A \dot A} H^{A \dot A} -24 X^{ab} X_{ab} 
-12 W^{AB} Y_{AB} -12 W^{\dot A \dot B} Y_{\dot A \dot B} \nonumber \\
&+& 3 P \overline{P} + \frac{3}{2} H^{A \dot A} H_{A \dot A} 
-12 \Phi^{A \dot A}_{ab} \Phi_{A \dot A}^{ab} -12 U^{A \dot A} U_{A \dot A}
+ 16i \rho^a_A \Lambda^A_a \nonumber \\
&-& 16i \rho^a_{\dot A} \Lambda^{\dot A}_a
-48 \rho^A_a W_{AB}^{\ \ \ Ba} +48 \rho^{\dot A}_a 
W_{\dot A \dot B}^{\ \ \ \dot B a} +48i \rho^A_a \rho^{B a} W_{AB} \nonumber \\
&+& 48i \rho^{\dot A}_a \rho^{\dot B a} W_{\dot A \dot B}
+48 \rho^{A a} \rho^{\dot A}_a U_{A \dot A} -48i \rho^{A a} 
\nabla_{A \dot A} \rho^{\dot A}_a \nonumber \\
&+& 48i \rho^{\dot A a} \nabla_{A \dot A} \rho^A_a +96i \rho^A_a 
\Phi_{A \dot A}^{ab} \rho^{\dot A}_b \label{i}
\end{eqnarray}
which is equivalent to saying that $I$ is no longer an independent field. This 
constraint implies that only the longitudinal part of $H_m$ belongs to the 
nonlinear multiplet; its divergence lies in the original Weyl multiplet.
One has off-shell identities relating the covariant derivatives of $\rho^a_A$
and the other auxiliary fields \cite{muller89,moura031}.

Altogether, these component fields form then the "old minimal" ${\cal N}=2$ 
40+40 multiplet \cite{fv79}: 
\begin{equation}
e_{\mu}^m, \psi_\mu^{A a}, A_\mu, \Phi_\mu^{ab}, Y_{mn}, U_m, \Lambda_A^a, 
X_{ab}, H_m, P, \rho_A^a
\end{equation}
The final lagrangian of "old minimal" ${\cal N}=2$ supergravity may be written,
in superspace, as ($\epsilon$ is the chiral density) \cite{muller84}:
\begin{equation}
{\cal L}_{SG} = -\frac{3}{4 \kappa^2} \int \overline{\epsilon}
d^4 \overline{\theta} + \mathrm{h.c.} \label{actionpure2}
\end{equation}

\section{Leading string corrections to supergravity}
\setcounter{equation}{0}
\indent

For many different reasons, it is important to know the quantum corrections to 
supergravity originated from string theory. The fourth power of the Riemann 
tensor $\left( {\cal R}^4 \right)$ shows up in all effective actions for all 
string theories and M-theory, with nonzero coefficients \cite{gw86}. In this 
sense, it is the most general quantum correction to supergravity one has. 
Therefore, it is of interest to know it, and the supersymmetric invariants to 
which it belongs, better.

The four-dimensional supersymmetrization of ${\cal R}^4$ had never been worked
out. A term like that was first considered as a possible quantum correction to 
supergravity \cite{dks77}; its study is the purpose of this work.

\subsection{The ${\cal N}=1$ case}
\indent

The lagrangian we will be considering is
\begin{equation}
{\cal L}_{SG}+{\cal L}_{{\cal R}^4}=\frac{1}{2 \kappa^2} \int E\left( 
1+\alpha W^2\overline{W}^2\right) d^4\theta \label{action}
\end{equation}
$\alpha$ is a coupling constant of mass dimension -6. The $\alpha$ term 
represents the supersymmetrization of one combination of the fourth 
power of the Weyl tensor, more precisely \cite{moura02}
${\cal W}_+^2 {\cal W}_-^2$. 

To compute the field equations from this lagrangian, we need the constrained
variation of $W_{ABC}$ \cite{wz781}. We presented elsewhere \cite{m01} the 
details of this calculation and the final result for $\int \delta \left[ 
E\left( 1+\alpha W^2\overline{W}^2\right) \right] d^4\theta$, and we do not 
reproduce them here again. From this result, the $R, \overline{R}$ field 
equations are immediately read:
\begin{equation}
R=6\alpha \frac{\overline{W}^2 \nabla^2 W^2}{1-2\alpha W^2
\overline{W}^2}=6\alpha \overline{W}^2 \nabla^2 W^2+12\alpha^2
\overline{W}^4 W^2 \nabla^2 W^2  \label{r}
\end{equation}
From (\ref{wb203}), we can easily determine $\nabla^n G_n$. This way, auxiliary
fields belonging to the compensating chiral multiplet can be eliminated 
on-shell. This is not the case for the auxiliary fields which come from the 
Weyl multiplet ($A_m$), as we obtained \cite{m01} a complicated differential 
field equation for $G_m$. 

\subsection{The ${\cal N}=2$ case}
\indent

To (\ref{actionpure2}) we are adding a correction given by
\begin{equation}
{\cal L}_{{\cal R}^4}= \alpha \kappa^4 \int \epsilon \phi d^4 \theta + 
\mathrm{h.c.} \label{action2chiral}
\end{equation}
where we have defined the chiral superfield ($W^2\overline{W}^2$ is acted by
the chiral projector)
\begin{equation}
\phi=\left(\nabla^{\dot Aa} \nabla_{\dot A}^b \left(\nabla_a^{\dot B} 
\nabla_{\dot Bb} +16 X_{ab} \right)  
-\nabla^{\dot Aa} \nabla_a^{\dot B} \left(\nabla_{\dot A}^b \nabla_{\dot Bb} 
-16i Y_{\dot A \dot B} \right) \right) W^2\overline{W}^2 \label{defphi}
\end{equation}
$\alpha$ is now a numerical constant. The term $\left. \epsilon 
\nabla^{Aa} \nabla^{Bb} \left[\nabla_{Ab}, \nabla_{Ba} \right]\phi \right|  
+\mathrm{h.c.}$ contains $e {\cal W}_+^2 {\cal W}_-^2$.

We then proceeded with the calculation of the components of $\phi$ and analysis
of its field content \cite{moura031}. For that, we used the differential 
constraints from the 
solution to the Bianchi identities and the commutation relations. The process 
is straightforward but lengthy. The results can be summarized 
as follows: with the correction (\ref{action2chiral}), auxiliary fields 
$X_{ab}$, $\Lambda_{\dot C c}$, $Y_{\dot A \dot B}$, $U_m$ and $\Phi_m^{ab}$ 
get derivatives, and the same should be true for their field equations; 
therefore, these superfields cannot be eliminated on-shell. We also fully 
checked that superfields $P$ and $H_m$ do not get derivatives with this 
correction (with the 
important exception of $\nabla^m H_m$) and, therefore, have algebraic field 
equations which should allow for their elimination on shell. The only obscure 
case is the auxiliary field $\rho_A^a$. We did not analyze its derivatives 
because that would require computing a big number of terms and, for each term,
a huge number of different contributions. This is probably because $\rho_A^a$ 
belongs to a nonlinear multiplet. We believe that its derivatives should 
cancel, though, because derivatives of the other fields from the nonlinear 
multiplet cancel.

\subsection{Conclusions and a conjecture}
\indent

A careful analysis shows that, in the cases we studied, the auxiliary fields 
that can be eliminated come from multiplets which, on-shell, have no physical 
fields; while the auxiliary fields that get derivatives come from multiplets 
with physical fiels on-shell (the graviton, the gravitino(s) and, in 
${\cal N}=2$, the vector). Our general conjecture for ${\cal R}^4$ 
supergravity, which is fully confirmed in the "old minimal" ${\cal N}=1$ case, 
can now be stated: the auxiliary fields which come from multiplets with 
on-shell physical fields cannot be eliminated, but the ones that come from 
compensating multiplets that, on shell, have no physical fields, can.
This analysis should also be extended to the other different versions of 
these supergravity theories.

\section*{Acknowledgments}
We gratefully acknowledge support from the Calouste Gulbenkian 
Foundation (through a short-term fellowship) and NSF (through grant 
PHY-0098527).

\end{document}